# Modal Analysis of Cellular Dynamics in the Morphospace in Epithelial-Mesenchymal Transition

Akash Chandra Das[1,$], Debanga Raj Neog[2], Biplab Bose[1,*]

[1]*Department of Biosciences and Bioengineering, Indian Institute of Technology Guwahati, Guwahati, India 781039*

[2]*Mehta Family School of Data Science and Artificial Intelligence, Indian Institute of Technology Guwahati, Guwahati, India 781039*

[$]*Current Affiliation: Department of Population Health Sciences, Weill Cornell Graduate School of Medical Sciences, Cornell University, New York City, USA 10065*

[*]*E-mail: biplabbose@iitg.ac.in*

**Abstract**

During epithelial-mesenchymal transition (EMT), epithelial cells change their morphology, disperse, and gain mesenchymal-like characteristics. Usually, cells are categorized into discrete cell types or states based on gene expression and other cellular features. Subsequently, EMT is investigated as a dynamical process where cells jump from one discrete state to another. In the current work, we moved away from this idea of discrete state transition and investigated EMT dynamics in a continuous phenotypic space. We used morphology to define the phenotype of a cell. We used the data from quantitative image analysis of MDA-MB-468 cells undergoing EGF-induced EMT. We defined the morphological state space or 'morphospace' using the morphological features extracted through image analysis. During EMT, as the morphology changed, the distribution of cells in the morphospace also changed. However, this morphospace had a very high dimension. We reduced it to a 2-dimensional "reduced morphospace" and investigated the temporal change in the spatial distribution of cells in this reduced space. We used proper orthogonal decomposition to find dominant dynamical features of this spatio-temporal data. The modal analysis detected key features of EMT in this experimental system − reversible transition, distinct paths of phenotypic transition during induction and reversal of EMT, and enhanced diversity of cells during reversal of EMT. We also provide some intuitive physical meaning of the spatial modes and connect them to the key molecular event during EMT.

**Keywords:** epithelial to mesenchymal transition; morphology; morphospace; phenotypic state transition; proper orthogonal decomposition

# I Introduction

Phenotypic state transitions are transformative processes through which a cell alters its molecular characteristics, morphology and functions. In other words, a cell changes from one "cell type" or "cell state" to another. Different aspects of metazoan biology, from embryonic development to cancer cell metastasis, involve phenotypic state transitions of cells.

Epithelial to Mesenchymal Transition (EMT) is a well-investigated example of phenotypic state transition. Multiple rounds of EMT and MET (Mesenchymal to Epithelial Transition) are involved in several stages of embryonic development, including gastrulation and neural crest formation [1,2]. In cancer cell metastasis, epithelial cells undergo EMT, thereby gaining the features of mesenchymal cells [3,4]. These EMT-derived cells also acquire stem cell-like properties and drug resistance [4,5].

Epithelial cells are tightly packed adherent cells. They attach to the extracellular matrix and stay tightly bound with adjacent cells. During EMT, they lose contact with neighbouring cells and the matrix [6,7]. These cells acquire characteristics of mesenchymal cells. They become more migratory and invasive [6,7].

Simultaneously, the morphology of these cells also changes. Most epithelial cells have cobble or polygonal shape. However, they change to more elongated or circular shapes as they undergo EMT [1,8,9].

All these changes during EMT are regulated through intricate regulatory molecular networks [10–12]. Certain key molecular circuits, like coupled feedbacks involving SNAIL, ZEB, TWIST, miR-200, and miR-34, have been identified that regulate the changes in gene expression during EMT [13]. Induction of EMT is associated with a decrease in the expression of epithelial markers like E-cadherin and β-catenin and an increase in the expression of mesenchymal markers like N-cadherin and Vimentin [14].

Based on the expression level of these molecular markers, we can define two cell states – epithelial (E) and mesenchymal (M). During EMT, the cell jumps from the E state to the M state. The opposite happens during MET. However, several studies have shown that there can be more than two states, with one or more intermediate states between E and M [15–18]. These intermediate states are known as partial or hybrid EMT states.

Whatever the number of cell states, most of these studies define cell states in terms of gene expression. However, the molecular markers are just proxies for cell phenotype. The phenotype of a cell is defined in terms of its physical and functional characteristics. The changes in the expression of EMT markers often vary considerably between different experimental systems [6,19,20]. However, physical characteristics like loss of adherence, change in shape, and gain of motility are universal for all cellular systems undergoing EMT. Therefore, it has been recommended that the EMT status be defined in terms of physical and functional features, like morphology and motility of a cell, rather than the level of molecular markers only [6].

Changes in the morphology of cells during EMT can be investigated using quantitive image analysis. Earlier, we investigated morphological changes during EGF-induced EMT in MDA-MB-468 cells, a breast cancer cell line [8]. We observed three predominant shapes for these cells – cobble, elongated spindle and circular. MDA-MB-468 cells are epithelial cells. These cells were tightly packed and mostly cobble-shaped. EGT treatment triggered EMT in these cells, turning them circular and spindle-shaped. We defined cell states in terms of cell morphology. We created a three-state cell state transition model and investigated the paths of state transitions using the data obtained through quantitative image analysis [8].

Like ours, other studies on phenotypic state transition consider discrete cell states. Categorizing cells in a finite number of discrete states allows the use of powerful mathematical formalisms, like the Markov Chain. However, one can argue that such discretization is a mere requirement of our mathematical analysis, and cell states are not inherently discrete. Phenotypic space may be continuous, with different regions occupied by clouds of similar cells, and due to plasticity, cells can move smoothly in this space [21–23].

Following the idea of a continuous state space, we reanalyzed the data from our previous study on EGF-induced EMT in MDA-MB-468 cells. This time, we do not categorize cells into discrete cell types. Rather, we consider a continuous multidimensional phenotypic space based on the morphological features of cells. We call it morphospace. A cell occupies a particular position in this space. During EMT, its morphology changes, moving it to another location in the morphospace.

Due to morphological diversity, an ensemble of cells creates cell 'clouds' in the morphospace. During EMT, the distribution of cells in the morphospace changes, thereby changing the shapes and positions of cell clouds. We follow the temporal dynamics of these cell clouds and investigate the underlying dynamical features through modal analysis. Our analysis successfully captures the key events and features of EMT in our experimental system.

## II Methods

**a) The data:**

We used the morphological data from our previous work on EGF-induced EMT in the breast cancer cell line MDA-MB-468 [8]. Details of the experiments are provided in the original article [8]. In the present work, we used the data from the experiments where cells were treated with 10 ng/ml of EGF for different durations till 60 hr. Subsequently, cells were fixed, stained with HCS cell mask red dye, and imaged using an Epi-fluorescence microscope.

Note that live-cell imaging was not used in these experiments. Therefore, we do not have time series data on morphological changes of individual cells. Rather, we had multiple samples (plates with thousands of cells) for each time point. Cells in those samples were fixed and imaged at a specific time. With thousands of cells in our images, we assume that the morphological dynamics has been adequately sampled in these experiments.

Post imaging, CellProfiler [24] was used for image segmentation, cell identification, and extraction of morphological features of each cell. We extracted a large number of geometric features (like eccentricity and form factor) and non-geometric features (like granularity and texture). Morphological features that showed sufficient variation with time (CV $\geq$ 0.1) were kept in the dataset.

**b) Dimension reduction and kernel density estimation:**

There were 85 morphological features in the final dataset. We used UMAP [25] to reduce the dimension of the data from 85 dimensions to two dimensions. Data of all time points were stacked in a single dataset before UMAP to maintain consistency. Post UMAP, data for each time point was extracted from this dimension-reduced data. On average, each time point had the data of ~12000 cells.

Subsequently, kernel density estimation [26] was used to find the distribution of cells in the 2-dimensional UMAP space at each time point.

**c) Proper orthogonal decomposition (POD):**

POD [27] was used to identify the dominant spatial modes in the dynamics of cell density in the 2D UMAP space. We used Singular Value Decomposition (SVD) for this purpose.

The 2D density data of each time point obtained by kernel density estimation was used. The kernel density data for each time point is a $120 \times 120$ matrix. Rows of this data were stacked to create a density vector **d** for each time point. We have data for 12 time points. These 12 density vectors were stacked side-by-side in the order of experimental time points. The stacking of vectors generated a matrix **D**. **D** is a $n^2 \times k$ matrix where $n = 120$ and $k = 12$.

This data matrix was centered for each row,

$$\mathbf{C} = \mathbf{D} - \bar{\mathbf{d}} \qquad (1)$$

$\bar{\mathbf{d}}$ is the matrix for row centering, and **C** is the centered data matrix. **C** is a $n^2 \times k$ matrix where $n = 120$ and $k = 12$. The $i$-th column of this matrix $c_i$ is the centered data of the $i$-th time point.

This matrix was decomposed using economy SVD,

$$\mathbf{C} = \mathbf{U}\mathbf{S}\mathbf{V}^T \qquad (2)$$

**U** is a $n^2 \times k$ matrix, and each column of **U** is a spatial mode. These modes are orthogonal to each other. **S** is a $k \times k$ diagonal matrix with $k$ singular values corresponding to each spatial mode. **V** is a $k \times k$ orthogonal matrix.

**d) Software:**

Data transformation, analysis, visualization, UMAP, kernel density estimation, and SVD were performed using MATLAB R2021a. SigmaPlot and MATLAB were used for data visualization.

## III Results

**a) EMT-associated morphological changes:**

A triple-negative adenocarcinoma cell line MDA-MB-468 was treated with EGF (10 ng/ml) to trigger EMT. Cells were treated for different durations – 0, 3, 6, 9, 12, 24, 27, 30, 33, 36, 48 and 60 h and imaged. These cells are adherent and largely of cobblestone shape. Upon treatment with EGF, cell-cell contacts loosened, and cells changed their morphology. Most cells turned to circular, elongated, and spindle shapes. These cells were still loosely adhered to the plate but were scattered individually or as small clusters of cells. Fig. 1 shows a few representative images at four time points.

The observed morphological changes are the signatures of EMT in this experimental system. Induction of EMT was also confirmed through molecular-marker-based analysis [8].

EGF-induced EMT in these cells was reversible [8]. Treatment with EGF induced a rapid change in the morphology and molecular profile of cells. However, with time, as the effect of EGF decayed, most cells returned to the initial state. As the initial changes were rapid, we imaged cells at an interval of 3 hr. From 24 hr, cells started reverting to the initial state. Imaging was performed at an interval of 3 hr to capture this reversal. Otherwise, imaging was performed at an interval of 12 hr.

**2) Dynamics of cells in the morphospace:**

A cell is a dynamical system with a state space. We extracted quantitative information on 85 morphological features for every cell through image analysis. So, we defined the cellular state space using these morphological features. We call this state space 'morphospace.' During EMT, with changes in morphology, cells move within this 85-dimensional morphospace.

The state space of a dynamical system has certain topological features like sinks and saddles. Sinks are stable steady states, and the system will reach a sink if it starts somewhere in its basin of attraction. A saddle separates two sinks. In a system with bifurcation, a change in one or more parameter values changes the state space's topology. Bifurcation leads to the disappearance of existing sinks or saddles and the appearance of new ones.

Imagine starting a dynamical system from thousands of initial positions in the state space. These thousands of copies of the system will evolve with time and eventually get distributed in the state space according to the position of sinks and saddles. Changes in the topology of the state space due to bifurcation will also change the system's distribution in the state space. Therefore, the positional distribution of a system in its state space informs us of the system's dynamics and can also be used to detect bifurcation. This approach has been used to generate pseudo-potential landscapes of genetic networks [28] and to understand cell-state transition dynamics [29].

We use a similar approach to understand the morphodynamics of cells. We investigate the temporal change in the distribution of cells in the morphospace. In other words, we investigate how the "cell cloud" moves in the morphospace.

In our experiments, samples for each time point were fixed, stained, and imaged. Consequently, we do not have time series data for individual cells. Therefore, we can not investigate the dynamics of individual cells in the morphospace. However, we imaged thousands of cells from multiple replicates at every time point. All these cells were from the same initial stock. These were seeded at the same time point and maintained in identical culture conditions. Therefore, we assumed that the distribution of cells in the morphospace at a particular time point in our experiments is similar to what we would have if we had followed the same set of cells over time using live cell imaging.

The morphospace is 85 dimensional. For dynamical analysis and to visualize the data, we reduced the dimension of the morphological data from 85 dimensions to 2-dimension using UMAP. Data for different time points, projected in the UMAP space, is shown in Fig. 2. We call this 2D UMAP space the "reduced morphospace".

To capture the temporal dynamics of the cell cloud, we estimated the distribution of cells in the UMAP space using kernel density estimation. The temporal changes in the estimated density are shown in Fig. 3.

The dynamics of cells in this reduced morphospace is quite evident in Fig. 2 and 3. The distribution of cells in this space changed with time. Initially, most cells are on the left-hand part of the space. On induction of EMT by EGF treatment, the cell cloud rapidly moved to the right-hand side.

The reversibility of EGF-induced EMT is also captured in this reduced space. After 12 hr, the cell cloud moved from the right-hand side to the lower left corner. Eventually, with time (from 33 hr), cells started moving towards the left-hand side of the plot that was initially occupied by the majority of cells.

**3) Modal decomposition of the cell cloud dynamics in the reduced morphospace:**

The estimated kernel density $g(\mathbf{x},t)$ is a function of the position on the reduced morphospace and time. We used proper orthogonal decomposition (POD) to decompose kernel density data into independent spatial and temporal components. The kernel density data of all the time points are stacked to create a composite, centered data matrix $\mathbf{C}$, where each column is the kernel density data for a time point. The details of this data reshaping and arrangement are explained in the methods section. $\mathbf{C}$ was decomposed using economy SVD,

$$\mathbf{C} = \mathbf{U}\mathbf{S}\mathbf{V}^T = \sum_{i=1}^{m} \sigma_i \mathbf{u}_i \mathbf{v}_i^T \tag{3}$$

The $i$-th-column of $\mathbf{U}$ is the $i$-th spatial mode $\mathbf{u}_i$. The $i$-th-column of $\mathbf{V}$ is the $i$-th temporal mode $\mathbf{v}_i$. $\sigma_i$ is the singular value for the $i$-th mode.

Each spatial mode ($\mathbf{u}$) captures a dominant pattern of spatial variation in the kernel density data. Singular values of these modes ($\sigma$) represent their relative importance, and the elements of respective temporal modes ($\mathbf{v}$) capture their time-varying contribution. Therefore, one can consider that the spatial modes are the hidden dynamical features in the spatio-temporal data.

We obtained $m = 11$ spatial modes with non-zero singular values. Each spatial mode is a $n^2$-by-1 vector, with $n = 120$. Each vector was reshaped into an $n \times n$ matrix corresponding to the 2D UMAP space (or reduced morphospace). These reshaped spatial modes are color-coded and are shown in Fig. 4.

Each element of a spatial mode corresponds to a particular spatial location on the reduced morphospace, and its numerical value represents its contribution to that spatial mode. In Fig. 4, these numerical values are color-coded. Regions with similar colours (say red) suggest regions with similar contributions to that mode. Therefore, regions with the same colour contribute similarly to the spatiotemporal dynamics on the reduced morphospace. Conversely, contrasting colours (say red and blue) indicate regions with opposite or anti-correlated contributions to the dynamics.

For example, the first spatial mode has two distinct regions, one in red and the other in blue. Therefore, these two regions are anti-correlated. When one of these regions has high cell density, the other will have very low cell density.

All spatial modes of a system do not contribute equally to the overall dynamics of the system. We select a few dominant modes in POD that can capture the system's dynamics. These dominant spatial modes are selected based on their singular values.

Fig. 5a shows the relative singular values of the spatial modes. The first four modes constitute ~73% of the cumulative singular values. These four spatial modes are dominant and can capture most of the dynamics of the cell cloud in the reduced morphospace.

We re-created the temporal behaviour of the cell cloud in the reduced morphospace using these four dominant modes,

$$\mathbf{D}_4 = \mathbf{U}_4 \mathbf{S}_4 \mathbf{V}_4^T + \overline{\mathbf{d}} \qquad (4)$$

$\mathbf{U}_4$ has the first four columns of $\mathbf{U}$. $\mathbf{V}_4$ has the first four columns of $\mathbf{V}$. $\mathbf{S}_4$ is a diagonal matrix with the four largest signal values in $\mathbf{S}$. $\overline{\mathbf{d}}$ is the row centering matrix used to generate the centered data $\mathbf{C}$ from original kernel density data $\mathbf{D}$.

$D_4$ is a $n^2 \times k$ matrix, where $n = 120$ and $k = 12$. Each column of $D_4$ represents the recreated kernel density data for a time point. Columns of $D_4$ are reshaped in $n \times n$ matrices (($n = 120$) and visualized in Fig. 5b. Visual comparison of Fig. 3 and 5b shows that the essential trends of the dynamics of cell density in the reduced morphospace are captured in this re-created data. That means we can investigate the dynamics of the cell cloud in reduced morphospace in terms of these four spatial modes.

**4) Morphodynamics during EMT and POD modes:**

In POD, we decompose the dynamics of a system into spatial and temporal modes. The $V$ matrix in Eq. (3) holds those temporal modes. The temporal contribution of each of the spatial modes is calculated using $S$ and $V$,

$$A = VS \qquad (5)$$

As we have 12 time points and 11 spatial modes, $A$ is a $12 \times 11$ matrix. The $i$-th column of $A$ represents the temporal contributions of the $i$-th spatial mode. We call the elements of $A$ temporal coefficients. Fig. 6a shows the time-dependent changes in the temporal coefficients of the four dominant modes.

The temporal coefficient varies around zero. If a spatial mode has a considerable time-dependent contribution to the system's dynamics, its temporal coefficient will vary considerably from zero.

The temporal coefficient of the most dominant spatial mode (mode 1) changes from a negative value to a positive one immediately after the EGF treatment and stays positive till 24 hr (Fig. 6a, upper panel). Subsequently, it reverts back to negative values. Therefore, spatial mode 1 contributes in two temporal phases. In one phase, its temporal coefficient is positive; in the other, it is negative.

In these two phases, spatial mode 1 has opposing effects on the overall dynamics of the cell cloud. Note that spatial mode 1 has two coherent zones with ani-correlation (red and blue coloured zones in Fig. 4). They represent two zones of high cell density in the reduced morphospace. One of these zones is dominant in the early phase of EMT when the temporal coefficient of mode 1 is positive. The other zone is dominant in the later phase when the temporal coefficient turns to negative values. The two anti-correlated zones on the reduced morphospace can be broadly considered as regions of two different cell types.

Fig. 3 shows that on induction of EMT, the cell cloud moved from the left-hand side of the reduced morphospace to the right-hand side, and then, slowly, with time, the cloud moved away from the right-hand side. This dominant left-right movement is captured by the spatial mode 1.

Our earlier work showed that the EGF-receptor (EGFR) phosphorylation is the primary regulator of morphological changes in EGF-induced EMT [8]. Fig. 6b shows the temporal changes in phospho-EGFR (pEGFR) in those experiments. Immediately after EGF treatment, the pEGFR level increased sharply, reaching a maximum by 12 hr. Subsequently, it dropped slowly to the basal level. The temporal behaviour of the pEGFR is similar to the temporal behaviour of spatial mode 1. In the period of high pEGFR, the temporal coefficient of spatial mode 1 is positive. As the phosphorylation of EGFR decays, the temporal coefficient switches back to negative values. The Pearson correlation coefficient between the average level of pEGFR and the temporal coefficient of mode 1 is 0.96 ($p = 0.002$).

These observations indicate that the dominant morphological changes during EGF-induced EMT are directly linked to the EGFR phosphorylation, and spatial mode 1 captures these dynamical changes.

Spatial mode 2 has distinctly different behaviour. Its temporal coefficient deviates from zero and turns to negative values after 24 hr, returning to low positive values by 36 hr. That means, in this window (24 to 36 hr), some unique morphological changes happened that are captured by the spatial mode 2.

EGF-induced EMT in our experimental system is reversible. Cells move back to the initial morphological state as the phospho-EGFR level decays. However, the morphodynamics during the reversal was distinct. As shown in Fig. 3, upon EGF treatment, the cell cloud accumulated on the right-hand side of the reduced morpho-space. However, as the pEGFR decays, the cell cloud peaks near the lower left-hand side before slowly moving towards the initial position. Therefore, cells transiently go through distinct morphological changes. This transient dynamic is captured by spatial mode 2.

The temporal coefficients of spatial modes 3 and 4 remain close to zero most of the time. That means these two represent the system's basal steady dynamical features and, if not much of our interest, to understand the morphodynamics.

The essential patterns of the temporal evolution of the cell cloud in the reduced morphospace are captured by spatial modes 1 and 2. Rather than investigating the movement of the cell cloud in the reduced morphospace, we can consider an ensemble of cells as a single unit or system and then investigate the movement of that system in the 2-dimensional space defined by spatial modes 1 and 2. We call this 2-dimensional space the "modal space".

This is achieved by projecting the kernel density data for each time point in the modal space,

$$\mathbf{P}_2 = \mathbf{D}^T \mathbf{U}_2 \qquad (6)$$

$\mathbf{U}_2$ has the first two columns of $\mathbf{U}$. $\mathbf{P}_2$ is the projected data matrix ($k \times k$). This projected data is visualized in Fig. 7a. Each circle in this figure represents the system's (or the cell ensemble's) position in the modal space at a particular time.

Initially, the system was at the left-hand side of the modal space. Immediately after induction of EMT, cells move to the right-hand side of this space. This indicates a rapid and drastic change in the morphology of most cells. It is like a discrete jump from one morphological state to another.

The system remains at the same location in the modal space from 3 hr to 12 hr. That means the morphology of cells in the ensemble remains largely unchanged during this period.

Subsequently, the system moves out of that position and moves towards the position at $t = 0$. However, the trajectory of reversal is quite different. Rather than moving directly following the line between $t = 3$ hr and $t = 0$ hr positions, the system traverses through a distinct and longer path. That means the morphodynamics of cells in the ensemble during the reversal of EMT is quite different from the one observed during the induction of EMT. This difference in the system's trajectory in the modal space during induction and reversal of EMT is equivalent to hysteresis in a dynamical system.

## 5) Connecting the continuous dynamics with discrete cell states :

In our earlier work, we classified cells into three morphological states or cell types – cobble, spindle and circular. We estimated the change in the distribution of these three cell types during EGF-induced EMT (Fig. 3d in [8]). We observed that most cells turned into circular morphology immediately after EMT induction. However, a mixed population of these three cell types was observed during the reversal. Eventually, by 60 hr, most cells were in the cobblestone state.

Can we connect the dynamics involving three cell states with the movement of cells in the 2-dimensional modal space? We segregated the UMAP data of all time points into three groups based on the cell types. Subsequently, kernel density estimation was performed in UMAP space for each cell type. The estimated densities are shown in Fig. 7b.

We projected kernel density data for each cell type on the modal space of spatial modes 1 and 2. The square boxes in Fig. 7c represent three cell types. We have overlaid the data of Fig. 7a in the same plot.

At $t = 0$, most cells were epithelial and of cobblestone shape. The projected position of cobblestone cells overlaps with the position of the cell ensemble at $t = 0$ in the modal space.

On induction of EMT, most cells turn circular. Accordingly, the projected position of circular cells overlaps with the position of the system on the modal space from 3 h to 12 h.

The reversal starts at 24 hr, and the system moves on a distinct trajectory. We observed an increase in elongated spindle cells during this period (Fig 3d in [8]). As shown in Fig. 7c, the spindle cell's projected position is near the plot's lower edge. The trajectory for the reversal moves towards it and then turns towards the initial position. Unlike the cobble and circular cells, the projected position of the elongated cells does not overlap with the trajectory of the cell ensemble. That means, while making the return journey, the cell population is a mixture of diverse morphologies, with some of those close to the spindle shape. By classifying cells only in three groups, we possibly underestimated the diversity in our earlier work.

## 6) Spatial modes and morphological features:

Spatial modes 1 and 2 captured the dominant morphodynamics. These modes were derived from the data on the reduced morphospace. We used UMAP for dimension reduction. It is not possible to identify the contribution of the original morphological features to UMAP axes. Therefore, we can not immediately connect the spatial modes with the original morphological features.

However, one may look for the correlation between the temporal behaviours of a spatial mode and a morphological feature. A high correlation will provide a physically intuitive meaning of a spatial mode.

We calculated the mean of 85 morphological features for all cells at each time point. Let $\bar{f}_i^t$ be the mean value of the $i$-the feature at time $t$. The temporal coefficient of the $j$-th mode at time $t$ be $v_j^t$. Vectors $\mathbf{f}_i$ and $\mathbf{v}_j$ holds the $\bar{f}_i^t$ and $v_j^t$ of all the time points. We calculated the Pearson Correlation ($r_{ij}$) between $\mathbf{f}_i$ and $\mathbf{v}_j$ for all morphological features and spatial modes 1 and 2. Any correlation that failed in the test of hypothesis ($p > 0.01$) was removed.

Spatial mode 1 has a high correlation with many morphological features. Thirty-two morphological features have $r \geq |0.95|$. Fig. 8 shows the temporal behaviour of the top six morphological features with very high correlations with spatial mode 1.

Spatial mode 1 captures the dominant pattern of variation in the distribution of cells in the reduced morphospace. This mode has two zones of coherence (shown in red and blue colours in Fig. 4). With the induction of EMT, the temporal coefficient of this mode flips to positive from negative (Fig. 6a), indicating a shift in the dominance of these two zones. In this way, mode 1 captures the key morphological event during EMT – the movement of cells from the left-hand side of the reduced morphospace to the right-hand side (Fig. 3). This morphodynamics corresponds to the emergence of circular cells (Fig. 7c).

MDA-MB-468 cells are largely cobblestone shaped. During EMT, they lose cell-cell and cell-matrix contacts and turn circular. Correspondingly, we see a drop in the values of three morphological features in the cell images – perimeter, maximum axis length (i.e. length of the major axis of an equivalent ellipse) and maximum ferret diameter (i.e. the largest possible diameter) (Fig. 8a).

Fig. 8b shows three features having the highest positive correlation with the temporal coefficients of mode 1. Those are form factor, contrast, and texture InfoMeas1. The form factor measures circularity and is 1 for a circle. Therefore, this feature increased towards 1, as the temporal coefficient of mode 1 flipped to positive at the early time points. InfoMeas1 is a Haralick texture feature. Contrast measures local variation in an image. We imaged cells after staining with a fluorescent dye. Circular cells were smaller in size, more uniform in shape, and had sharp edges in the images. Possibly for these reasons, their contrast and InforMeas1 values were high for circular cells.

After 12 hr, the phospho-EGFR level dropped, and cells changed their morphology. Consequently, the cell clouds moved in the reduced morphospace. This reversal is captured again by the spatial mode 1. This time, its temporal coefficients change from positive to negative values. Accordingly, the highly correlated morphological features also changed. The features that capture circularity dropped in their values, indicating that cells were acquiring other shapes.

Spatial mode 1 captures the left-right movement of the cell cloud on the reduced morphospace. The reverse journey, after 12 hr, followed a longer path. This diversion is captured by spatial mode 2. Its temporal coefficient changes the sign between 24 and 36 hr. However, the temporal coefficients of mode 2 do not correlate significantly with any morphological features. In the reverse journey, cells had a more diverse morphology. Possibly, this diversity can not be captured through individual morphological features. Unlike a circular cell, cells in this period might not have standard geometric shapes. Such irregular shapes can not be captured in terms of individual morphological features. Therefore, we failed to find any one-to-one correlation between these features and spatial mode 2.

## IV Discussion:

The morphology of a cell is a quantifiable direct phenotypic measure that can be used to study phenotypic changes during EMT [8,30,31]. For the convenience of analysis and mathematical modelling, we categorize cells into distinct, discrete morphological groups or cell types. However, in principle, morphological space is continuous, and the cells, even in a discrete group, can show a large diversity in shape. Discretization may cause a loss of information on this diversity.

In this work, we do not consider any discrete cell type. Rather, we are interested in the dynamics of cells in a continuous morphological state space. As explained earlier, the distribution of cells in the state space captures the state space topology. One expects to see a high density of cells around a sink or attractor rather than a saddle. Each sink could be considered as a stable phenotype. As the system undergoes bifurcation triggered by an EMT-inducing molecular signal, the state space morphs, existing sinks disappear, and new ones are formed. Therefore, the spatial distribution of cells in the state space changes over time.

In this work, we defined the state space in terms of the morphological features extracted from quantitative image analysis. It is 85 dimensional space. We reduced the dimension using UMAP and then calculated the spatial distribution of cells in the 2D reduced morphospace using kernel density estimation.

Rather than chasing individual cells in the morphospace, we investigated the dynamics of the cell density in the reduced morphospace. Therefore, our approach is suitable for both live-cell and fixed-cell imaging data.

We reduced the dimension of the morphospace massively. Therefore, we can not expect to capture all the structures in the original morphospace. Even then, the kernel density maps in the reduced morphospace show phenotypic changes (Fig. 3). Initially, the cell cloud was on the left-hand side of the reduced morphospace. On induction of EMT, it moved to the right-hand side. This shift is distinct and indicates a sharp change in cell phenotype. From 24 hr onwards, phospho-EGFR level dropped, and cells slowly reversed to the epithelial phenotype. During this journey, the cell cloud moved through the lower part of the reduced morphospace. Again, this movement shows the emergence of new phenotypes.

We used UMAP for dimension reduction. It is a manifold learning technique [32]. Manifold learning methods have been used to understand and visualize the differentiation of cells from single-cell gene expression data [33,34]. The key assumption of these studies is that even though the original gene expression space is of very high dimension (usually thousands of genes), the effective state space for the cellular dynamics is of lower dimension.

We also made a similar assumption. Though the original morphospace is 85-dimensional, the 2-dimensional reduced morphospace successfully captured the overall morphodynamics. However, we would like to remind the reader that currently, there is no a priori method to know if such an assumption is correct. Also, we do not have any estimate of the biologically relevant information we may have lost on dimension reduction. We were able to capture the morphodynamics in the reduced dimension, possibly because some morphological features are redundant and have inter-dependencies.

The temporal changes in the kernel density over the reduced morphospace hold the dynamical information of phenotypic transition during EMT. If we could find a governing equation for the movement of the cell cloud, we would be able to describe the EMT dynamics explicitly. However, deriving such a governing equation is difficult as we do not have any known physical principle for the morphodynamics during EMT. Note that the kernel density data at each time point is two-dimensional, with 120 values in each dimension. Even if we consider an oversimplified spatial model for the time evolution of this system, that would have 120×120 variables. We cannot fit such a higher dimensional model to our limited data.

To circumvent these limitations, we used POD. POD can provide valuable insights into the dynamics of a system, even when the governing equations are unknown or too complex to solve or fit data. It extracts a dynamical system's dominant patterns or modes from limited snapshots of its behaviour.

We observed that four dominant spatial modes are good enough to recreate the kernel density data (Fig. 5b). Out of these modes, the first two hold most of the information to understand the phenotypic dynamics during EMT. The dominant dynamic of cells in the reduced morphospace is left-right movement. Such movement happened during both the induction of EMT and its reversal. Spatial mode 1 represents this dominant dynamics.

The key regulatory event in EGF-induced EMT is the phosphorylation of EGFR. In our experiment, EGF induced a rapid, sharp increase in phospho-EGFR (Fig. 6b). Phosphorylation of this receptor triggers many signalling cascades that reduce cell-cell and cell-matrix adhesion [35]. It also triggers cytoskeletal structures [36]. Therefore, the morphology of cells changes. In our experiment, cells turned roughly circular, which happened very fast, as did the EGFR phosphorylation.

Temporal coefficients are the measures of the contribution of a spatial mode to the system behaviour at different time points. The phospho-EGFR dynamics is strongly correlated with the temporal coefficients of spatial mode 1 (Fig. 6). Therefore, we can say that spatial mode 1 represents the dominant morphodynamics induced by EGFR-phosphorylation. In fact, we observed that several morphological features that can capture the circular cell shape had a very high correlation with the temporal coefficients of mode 1 (Fig. 8).

Spatial mode 1 has two distinct zones that are anti-correlated (Fig. 4). These two represent two dominant cell morphologies in our experiment. During the fast induction phase of EMT, cells moved from one dominant zone to the other. However, the decay of pEGFR was slow. Accordingly, the reversal of EMT was also slower. Cells with other diverse morphologies appeared during this reverse journey. This dynamics is captured by the spatial mode 2. This diversity in cell morphology was dominant during between 24 and 36 hr.

The morphological changes and associated movement of cells in the morphospace are distinct for the EMT induction and reversal phase. This difference is more evident when we project the ensemble of cells on the modal space (Fig. 7a). One could consider this as the hysteresis in EMT dynamics. A dynamical system with bifurcation, particularly bistable or multi-stable systems, shows hysteresis. Bistable and tri-stable molecular networks are known in EMT [13].

Can we say the difference in the paths on the modal space is the signature of bifurcation? We think that will be an over-assertion. Bifurcation is reported for the molecular networks of EMT [13]. In those studies, investigations were done on the molecular state space of a cell. A molecular state space is distinct from the phenotypic state space (in this case, morphospace). We can not expect a bijective relation between these two different spaces. Therefore, the morphospace may not reflect the multistabilities and bifurcations observed in molecular state space.

One can provide an alternate explanation for the observed path difference. The induction of EMT was fast, with a sharp increase in pEGFR. However, the reversal was slower. Due to the fast transition, cells moved almost discreetly from the left side of the reduced morphospace to the right side. We did not observe a diverse cell population because the dynamics was faster than our sampling intervals. Those missed morphologies were captured in the reverse phase as the dynamics were slow.

Here, we like to point out that POD can detect bifurcation [37,38]. We can define the system's entropy using singular values of the modes, and the entropy shows a distinct jump due to bifurcation. However, such an analysis requires adequate snapshots before and after the bifurcation. We could not perform such an analysis due to the limited snapshots.

Our previous study [8] also identified different paths for cell state transition during the induction and reversal of EMT. We considered three distinct cell types: cobble, elongated, and circular. Our discrete cell state transition model predicted that during the induction phase, the dominant state transition was Cobble → Circular. In contrast, the dominant path during the reversal of EMT was Circular → Elongated → Cobble.

Though the discrete state transition model and the modal analysis successfully identified the path difference, the modal analysis has certain advantages. Modal analysis is assumption-free – we did not assume a specific number of cell states. Further, it does not require parameter estimation by fitting the model to the data. The system's behaviour on the modal space clearly indicated the existence of two distinct paths of morphodynamics.

One can argue that the discrete cell state model has the advantage that the defined cell types are physically meaningful. Modal analysis on the morphospace does not immediately inform us of the real morphology of cells. Though this argument is valid, we observed that the most dominant spatial mode corresponds with measured, physically meaningful morphological features (Fig. 8). Further, the three morphological states of the discrete state transition model were mapped correctly on the modal space (Fig. 7c).

Interestingly, we could observe that the path on the modal space during EMT reversal does not touch the elongated cells mapped on this space (Fig. 7c). That shows that cells during this transition had much more diverse morphology. By considering just three states, we failed to capture that diversity in cell morphology.

Discretization had another problem. It made us believe that the population distribution of cells at 60 hr was close to that in the beginning. However, the trajectory on the modal plane shows that the system's position at 60 hr differed from that at 0 hr. That shows the diversity in morphology before induction of EMT was not the same when it was reversed. Possibly, cells needed more time to return to the original morphology.

We think the modal analysis of the morphospace complements discrete cell state-based modelling. One can also use this technique for problems where the state space is defined in terms of other quantitative phenotypic features.

Figure 1.

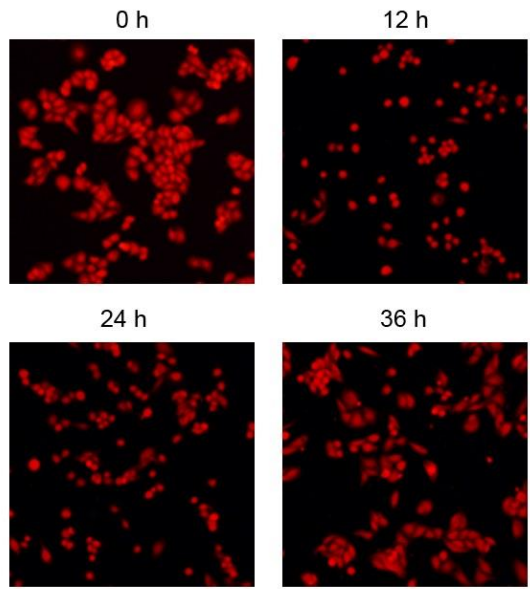

Figure 1. Morphological changes during EGF-induced EMT. MDA-MB-468 cells were treated with EGF (10 ng/ml) for different durations. Subsequently, cells were fixed, stained with a fluorescent dye, and imaged using an epi-fluorescence microscope.

Figure 2.

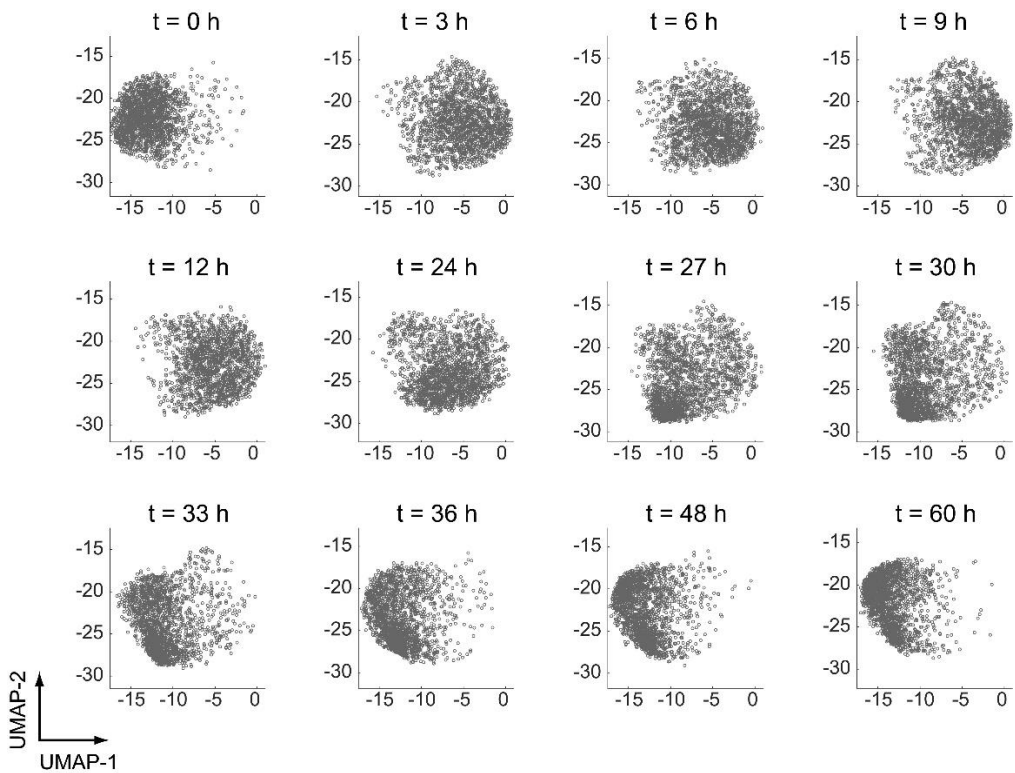

Figure 2. Dynamics during EGF-induced EMT in the reduced morphospace. The dimension of morphological feature data was reduced by UMAP from 85 dimensions to two. Each dot represents one cell. On average, ~12000 cells were imaged for each time point. However, for clarity in the plots, data of randomly selected 2000 cells are shown for each time point. Time points post-EGF treatment are written above each figure.

Figure 3.

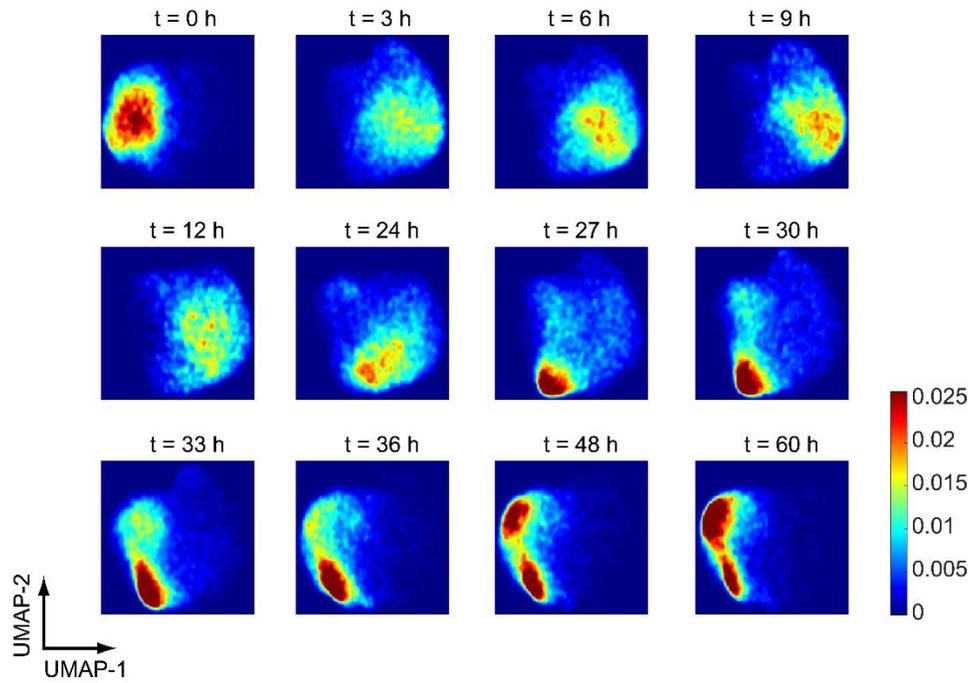

Figure 3. Temporal changes in the distribution of cells in the reduced morphospace during EGF-induced EMT. Data of kernel density estimation is shown for each time point. The colour bar shows the estimated density. Red means high, and blue means low.

Figure 4.

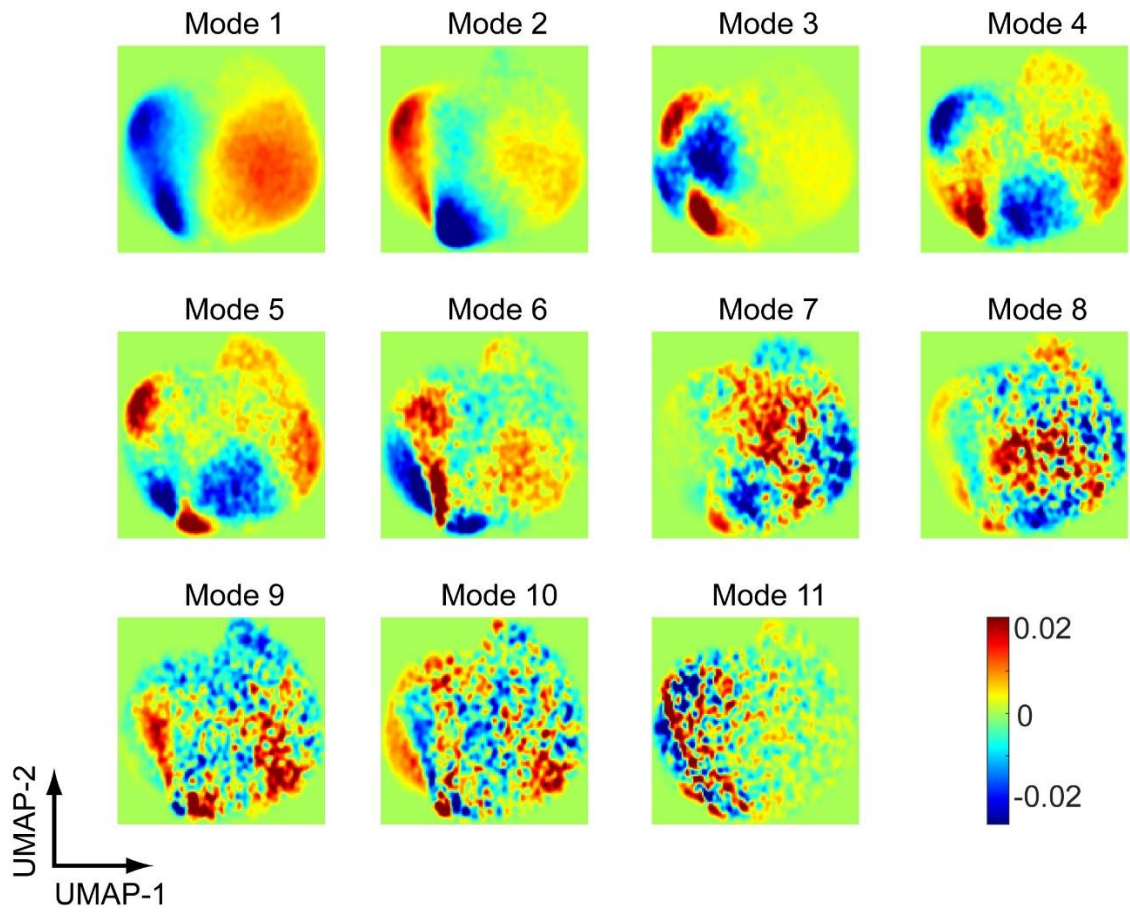

Figure 4. The spatial modes of the temporal kernel density data. The time-dependent kernel density data was decomposed using POD. Each 1D spatial mode vector was reshaped into 2D and visualized in the original UMAP space. Numerical values of the elements of each vector are colour-coded according to the colour bar shown.

Figure 5.

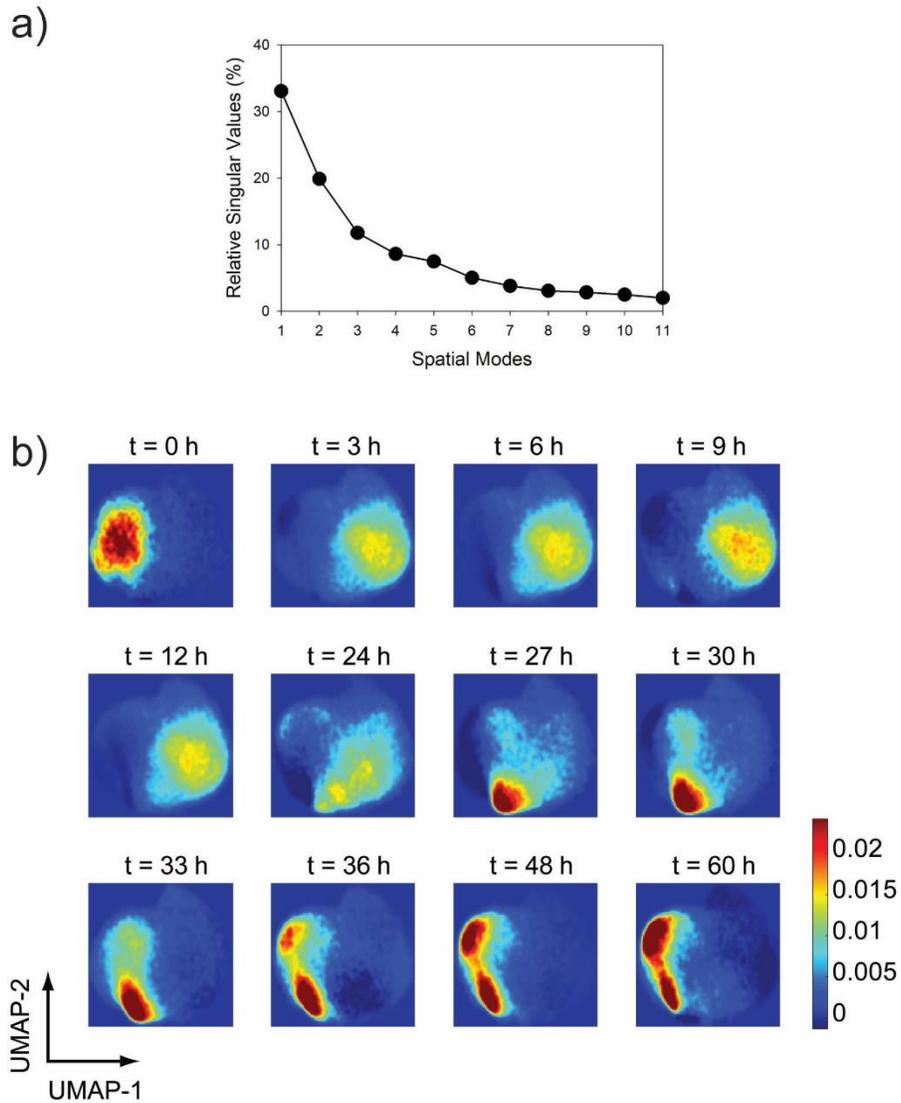

Figure 5. Four dominant spatial modes can recreate the dynamics in the reduced morphospace. a) Relative singular values of the spatial modes are shown. Relative singular value of the *i*-th mode $s_i = (\sigma_i \times 100)/\sum_{i=1}^{11}\sigma_i$, where $\sigma_i$ is the *i*-the singular value. b) Shows the Kernel density data for each time point, recreated using the first four spatial modes.

Figure 6.

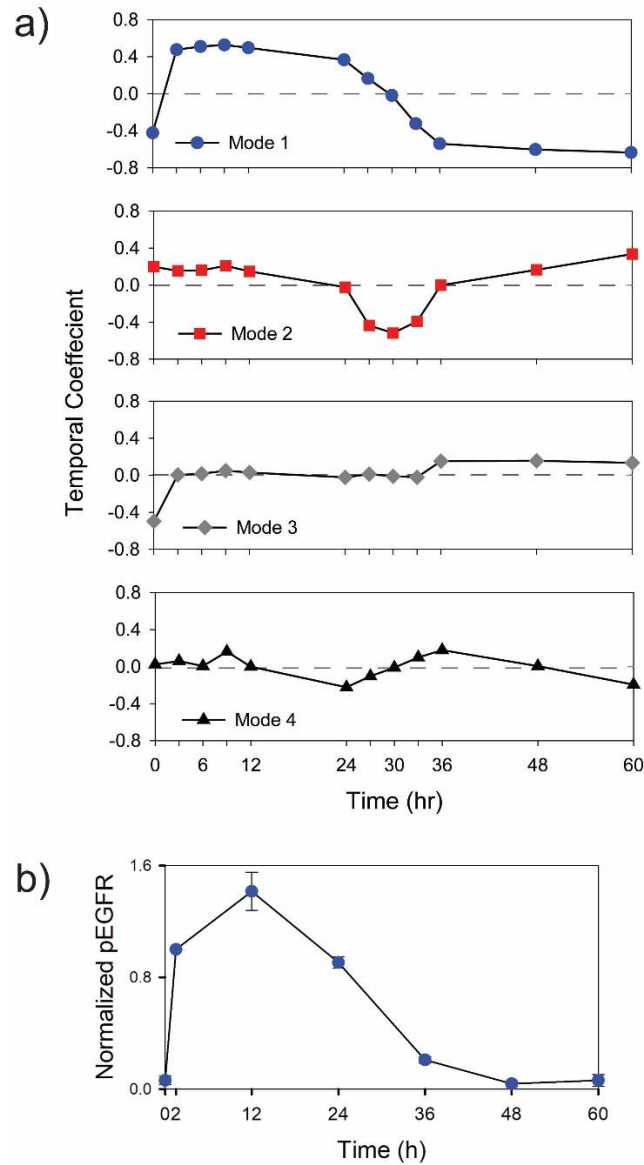

Figure 6. Temporal behaviour of four dominant spatial modes. a) Time-dependent behaviours of the temporal coefficients of the first four dominant spatial modes are shown. b) Normalized level of phosphorylated EGFR (pEGFR) at different times. This data was generated from the densitometric analysis of Western blots described in our earlier work [8]. The Western blot data of total EGFR in the same samples were used for normalization.

Figure 7.

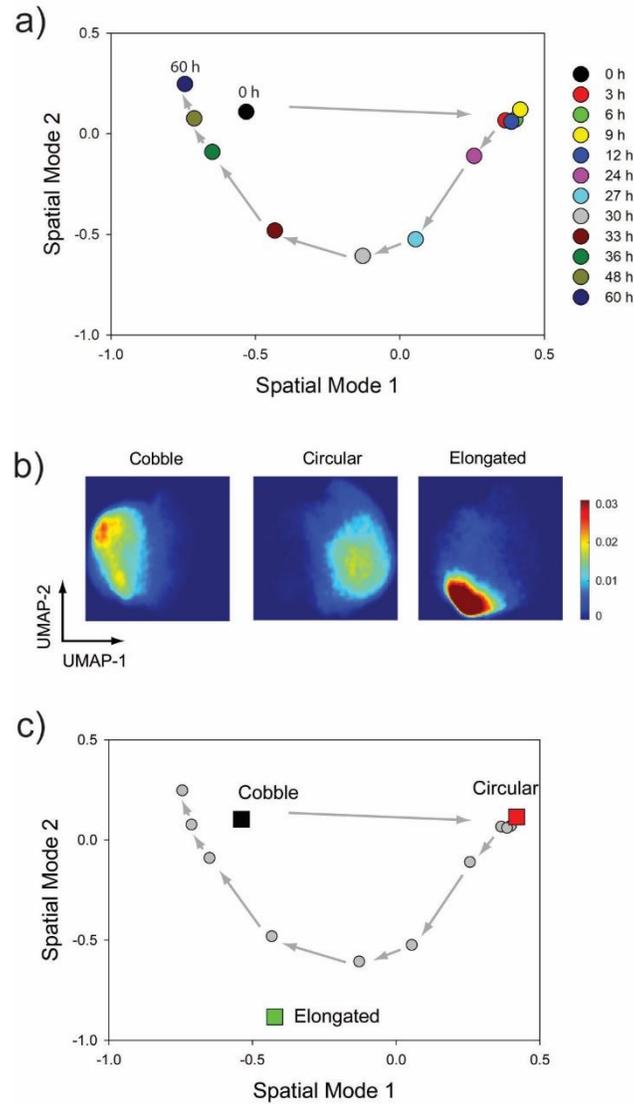

Figure 7. Dynamics of the cell ensemble in the modal space. a) Kernel density data of each time point projected on the modal space. Each circle represent the position of the cell ensemble at a particular time. Circles are colour-coded for time, and the arrows show the progression of time. b) Cells were categorized into three discrete types: cobble, circular and elongated. The distributions (or kernel densities) of these cell types in the reduced morphospace over all the samples are shown. c) The kernel density data of each cell type was projected in the modal space and overlayed on the plot shown in (a). Square boxes are cell types.

Figure 8.

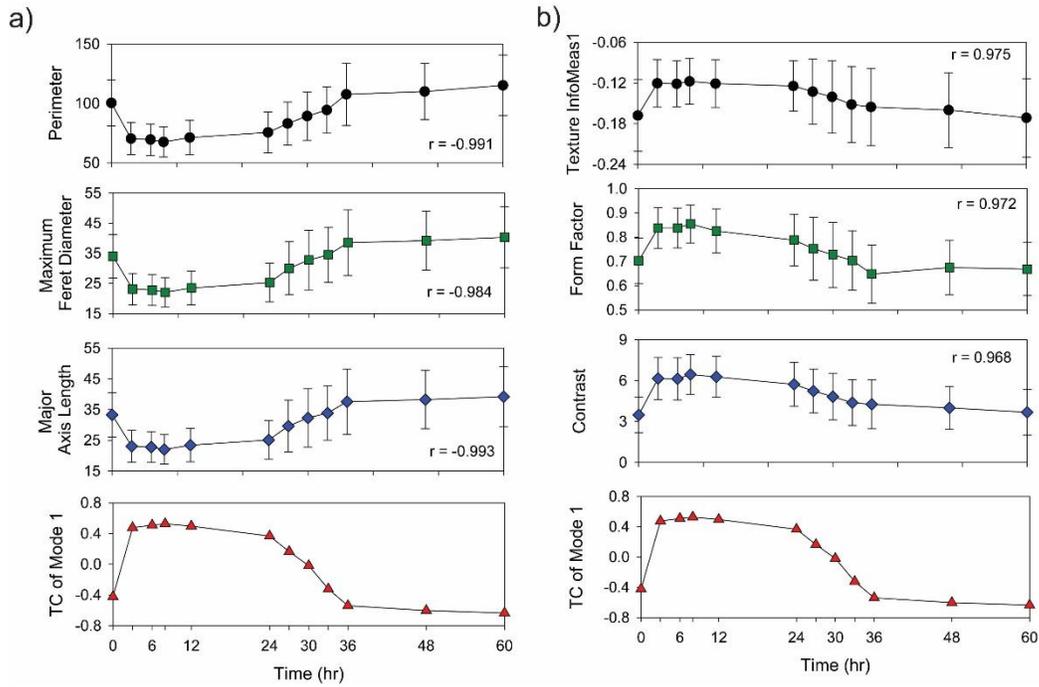

Figure 8. Temporal behaviour of selected morphological features with high correlation with spatial mode 1. a) shows three features with the highest negative correlation and b) shows those with positive correlation. The Pearson correlation coefficient for each case is written on respective plots. Points represent the mean values, and error bars show the standard deviations. The lower plots in both show the temporal coefficients (TC) of mode 1 over time.